# Redefining the limit dates for the Maunder Minimum

J. M. Vaquero[1,2] and R. M. Trigo[2,3]

[1]Departamento de Física, Universidad de Extremadura, Avda. Santa Teresa de Jornet, 38, 06800 Mérida (Badajoz), Spain (e-mail: jvaquero@unex.es)

[2]CGUL-IDL, Universidade de Lisboa, Lisbon, Portugal

[3]Departamento de Eng. Civil da Universidade Lusófona, Lisbon, Portugal

Abstract. The Maunder Minimum corresponds to a prolonged minimum of solar activity a phenomenon that is of particular interest to many branches of natural and social sciences commonly considered to extend from 1645 until 1715. However, our knowledge of past solar activity has improved significantly in recent years and, thus, more precise dates for the onset and termination of this particularly episode of our Sun can be established. Based on the simultaneous analysis of distinct proxies we propose a redefinition of the Maunder Minimum period with the core "Deep Maunder Minimum" spanning from 1645 to 1700 (that corresponds to the Grand Minimum state) and a wider "Extended Maunder Minimum" for the longer period 1618-1723 that includes the transition periods.

Gustav Spörer was the first to show in 1887 that there was a period of very low solar activity that ended in 1716. However the confirmation on the prolonged hiatus of solar activity would be attributed to the English astronomer Edward Walter Maunder who provided evidence that very few sunspots had been observed between 1645 and 1715 (Maunder 1922). John Allen Eddy, who popularized the name of this period as "Maunder Minimum" (MM), corroborated these dates after a thorough analysis of all available records of solar activity proxies including records of sunspot and aurorae observations from the 17th- and 18th-century astronomers and carbon-14 record (Eddy 1976). Since then, there has been a general consensus within the wider scientific community (from astrophysicists to historians) to use as limiting dates of MM the years 1645-1715, i.e. a 70 year long length. From cosmogenic isotope records, we know there are normal periods of solar activity intermingled with periods of unusually high or low activity, coined Grand Minimum or Maximum episodes respectively. The total duration of Grand Minima presents a bimodal distribution, with a peak of short durations of 30-90 years (with maximum distribution around 50-60 years) while others have long durations (>100 years) (Usoskin et al. 2007; Usoskin 2013). Currently, the best known Grand Minimum is the MM since we have direct observations of the Sun activity through the use of the telescope (Hoyt and Schatten 1998, Soon and Yaskell 2003).

However, our knowledge of past solar activity has improved significantly in recent years, underscoring a slightly different picture. A more comprehensive assessment of the onset, maturation and termination of the entire MM period can be appreciated through the use of different indices related to complementary aspects of this Grand Minimum. Some of these are shown in Figure 1 that depicts three distinct solar activity indexes for the period 1550-1800 including (green) decadal sunspot numbers, reconstructed by Usoskin et al. (2014), (orange) number of auroral nights per year in Hungary (Rethly and Berkes, 1963), and (blue) Group Sunspot Number (Hoyt and Schatten 1998). We must stress that the widely used GSN series developed by Hoyt and Schatten presents caveats and weaknesses and several corrections are being incorporated. In particular the GSN was improved for various erroneous years close to the edges of the MM period, including 1637-40, 1642, and 1652 (Vaquero et al. 2011, Vaquero and Trigo 2013).

We would like to note some interesting aspect of the solar activity indexes represented in Figure 1. Decadal sunspot numbers reconstructed by Usoskin et al. (2014) result from a physical reconstruction of solar activity using the latest verified carbon cycle, $^{14}$C production, and archeomagnetic field models. Therefore, this series provides an unprecedented level of detail at the decadal time-scale.

We have also used a regional auroral series developed in Hungary. This choice, instead of the widely used global auroral observations dataset by Křivský and Pejml (1988), can be puzzling at first glance. However, the huge data-set assembled by Křivský and Pejml (1988) presents a serious problem of inhomogeneity just after MM. In general, the compilers of these auroral catalogues used non-scientific sources before MM. However, with the recovery of the level of solar activity after the MM and from the famous aurora 1716, many scientists began to record their observations of auroras. Therefore, the total number of observations presents a ten-fold increase due to this fact (see Figures 6.24 and 6,25 in Vaquero and Vázquez 2009). This fact largely precludes the use of such records to compare the auroral activity before and after the MM. Some authors have made some attempts to correct for this effect, such as the use of "civilization factors" suggested by Křivský (1984). However, this procedure raises additional concerns as it represents a rather artificial fix to the problem. Interestingly, the series of auroras observed in Hungary were built using non-scientific literature and appear to insure better uniformity and a more straightforward link to solar activity. Figure 2 presents a direct comparison of auroral series provided by Rethly and Berkes (1963) and Křivský and Pejml (1988) showing the facts cited here.

Finally we would like to make a comment on the historical counting of sunspots. We can use an additional (and rather simple) index based in this counting: the annual percentage of active days, i.e., days with sunspot activity. Figure 2 shows this index from 1610 to 1800 using a 5-year moving average. The values of this simple index indicate a common behaviour respect to the other solar activity indices. In addition, the values of this index during the periods 1618-1645 and 1700-1723 are comparable to the values in the Dalton Minimum (1795-1820 approximately) which is not considered a Grand Maximum.

Overall it is possible to state that the three indices represented in Figure 1 agree in the identification of a Grand Minimum centred in the mid of the second half of the 17th-century. However, we would like to argue that the one-fits-all definition for the MM limits is not well supported by this multi-index evidence, where two transition periods are clearly visible in Figure 1. Therefore, our proposal here is two define two contiguous regions, 1) the first one characterised by virtually no solar activity and coined by Usoskin (2013) and others as "Deep Maunder Minimum" (DMM), spanning from 1645 to 1700 and 2) a wider "Extended Maunder Minimum" (EMM) with some solar activity and covering the much longer period 1618-1723. In any case, we must stress that the Grand Minimum period corresponds to the DMM with a duration of 55 years, just in the maximum of the bimodal distribution of durations provided by Usoskin et al. (2007).

Thus, in our view, the Maunder Minimum scenario can be summarized as follows: (i) two solar cycles of small amplitude in the period 1618-1645 and associated with strong decline in the decadal sunspot numbers reconstructed by Usoskin et al. (2014) and frequency of auroral nights, (ii) a period of deep solar activity minimum (1645-1700) where the 11-year cycle can not be clearly seen in the GSN (but still present at low levels in other proxies), and (iii) two solar cycles of small amplitude in the period 1701-1723 with clear increasing of signal in the other proxies. Therefore, the EMM would be formed by a "decay phase" (1618-1645), a "Grand Minimum" phase (1645-1700) and a "recovery phase" (1700-1723).

Although we know that the limit dates of MM could vary slightly if we use different criteria or different indices, we are confident that it is important to provide to the scientific community more precise dates for the onset and termination of a phenomenon that is of particular interest to many branches of natural and social sciences today (Soon and Yaskell 2003). In fact, recent studies have shown that a number of climatic and social anomalies occurred just during the two solar cycles of small amplitude previous to the MM (Uberoi 2012; Parker 2013). Although the magnitude of the change between the Total Solar Irradiance during the MM and the present day is uncertain, it is clear that the radiative output of the Sun was reduced, leading to a colder climate in general (Gray et al 2010). These changes in the solar output raise a clear need to constrain the entire MM period within more precise limiting dates.

## Acknowledgements

Support from the Junta de Extremadura (Research Group Grant No. GR10131), Ministerio de Economía y Competitividad of the Spanish Government (AYA2011-25945) and the COST Action ES1005 TOSCA (http://www.tosca-cost.eu) is gratefully acknowledged. We would also like to thank the anonymous reviewer for his constructive comments.

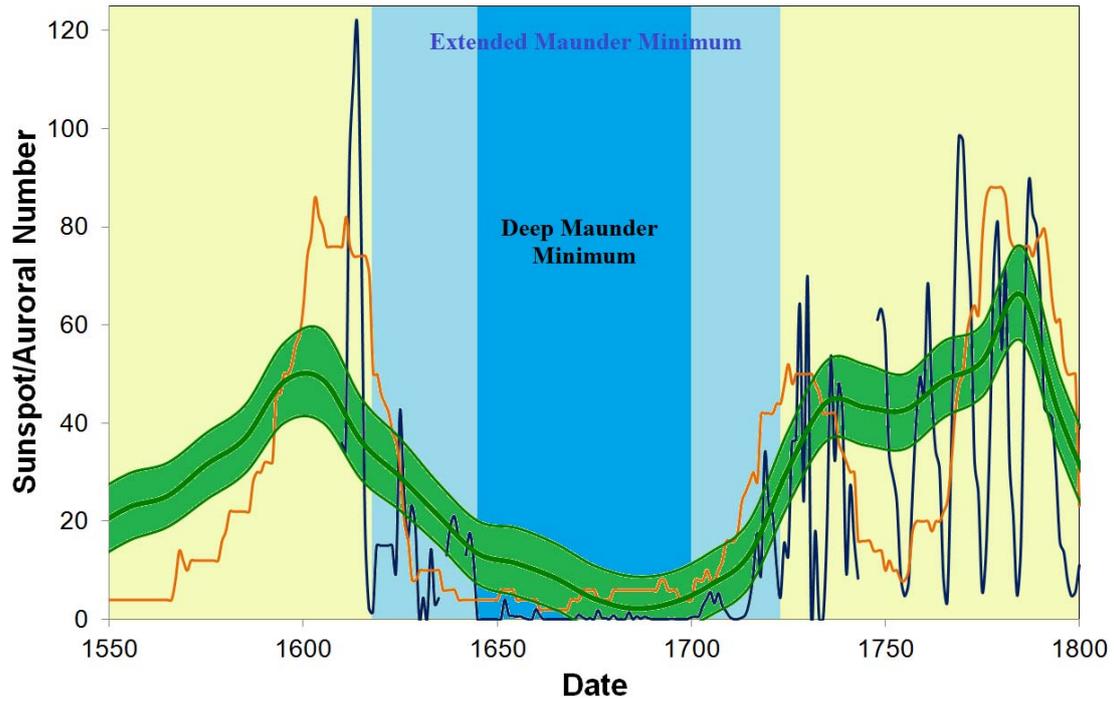

Figure 1. Solar activity indexes for the period 1550-1800: (green) decadal sunspot numbers, reconstructed by Usoskin et al. (2014), (orange) 25-year moving average of auroral nights in Hungary (Rethly and Berkes 1963), and (blue) Group Sunspot Number (Hoyt and Schatten 1998, Vaquero et al. 2011).

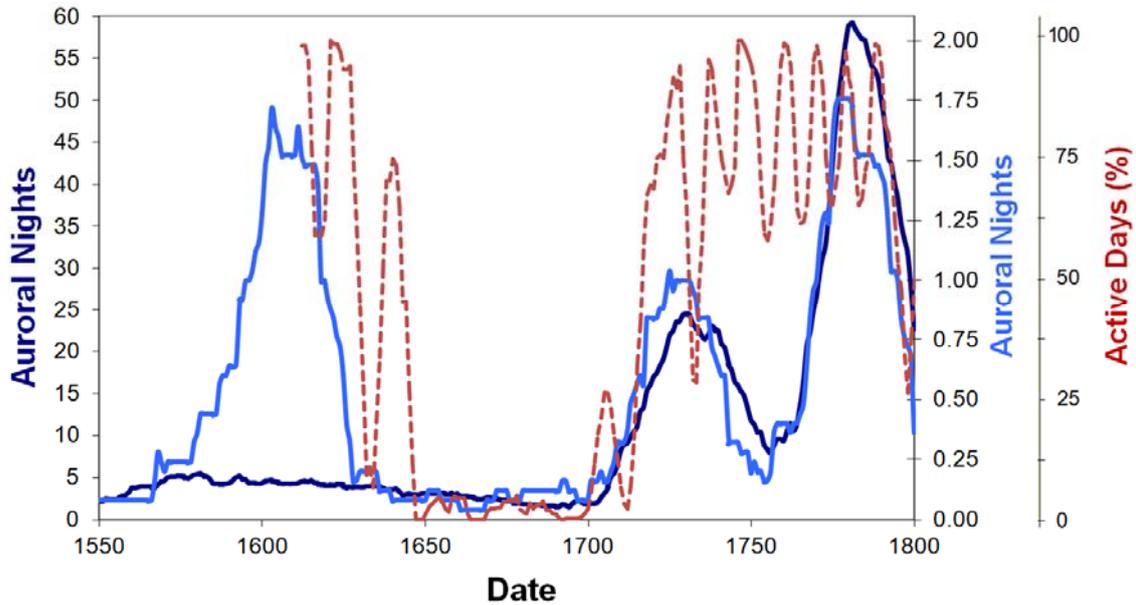

Figure 2. Auroral nights per year from 1550 to 1800 from (light blue) Rethly and Berkes (1963) and (dark blue) Křivský and Pejml (1988) using a 25-year moving average and (dashed red line) percentage of active days from 1610 to 1800 from Group Sunspot Number series using a 5-year moving average.